# Comment on LiH as a Li$^+$ and H$^-$ ion provider

**E.S. Skordas**

Department of Solid State Physics, Faculty of Physics, University of Athens, Panepistimiopolis, 157 84 Zografos, Greece

**Abstract**

A first-principle study of the formation and migration of native defects in LiH, a material of interest in hydrogen storage and lithium-ion batteries, has recently been published. Their results are found here to be of key-importance to deduce useful results for the pressure dependence of the ionic conductivity of LiH.

## 1. Introduction

In view of its very large hydrogen content, LiH is in principle of great interest for hydrogen storage [1], but its relatively high decomposition temperature [2] (~720°C) limits its usefulness for practical applications. However, upon combining LiH (which is also present in lithium-ion battery electrodes [3]) with other complex metal hydrides, experimental studies has shown that the dehydrogenation temperature is lowered and the hydrogen kinetics in improved [4, 5], thus partially overcoming the difficulties for practical applications. A better understanding of such phenomena requires a detailed knowledge of the formation and migration of native point defects in LiH. Along these lines Hoang and Van de Walle [6] (HVW hereafter) just published a first-principle study of these defects in which the energy calculations were based on density functional theory (DFT) [7] within the Perdew-Burke-Ernzerhof version [8] of the generalized-gradient approximation (CGA) and the projector augmented wave method [9, 10] as implemented in the VASP code [11-13]. Further the Heyd-Scuseria-Ernzerhof hybrid functional [14, 15], hereafter labeled HSE06, was also used in some bulk and defect calculations. In their calculation, HVW implemented finite-cell-size corrections in the Freysoldt sheme [16]. This requires values for the static dielectric constant for which HVW found for the electronic contribution the values 4.32 in CGA and 3.64 in HSE06, while for the ionic contribution the value 10.65 (obtained in CGA) was used for both CGA and HSE06 calculations. Thus, the calculated HVW values for the static dielectric constant [17-19] were 14.97 in CGA and 14.29 in HSE06, which exceed by only ~10% the corresponding experimental value.

The aforementioned detailed calculations of HVW showed that the negatively charged lithium vacancy and positively charged hydrogen vacancy are the dominant defects. Further, HVW found that ionic conduction occurs via a vacancy mechanism with an activation energy of about 1.37 eV which is somewhat lower than the experimental value [19] of 1.695±0.005 eV. As for the calculated migration barrier for the current-carrying defect is 0.51 eV which is in excellent agreement with the experimental value 0.53±0.02 eV for the lithium vacancy motion.

It is the aim of this short paper to discuss further the HVW first-principles calculations and show that their results are of wider usefulness if we complement them with the conclusions drawn on the basis of a macroscopic model based on thermodynamics which is termed cBΩ model [21-26] (see below). This model, which has led to successful results in a variety of solids including metals, ionic solids, rare gas solids and diamond, has been also recently applied to LiH as well as to silver-halide-cadmium halide systems [27] and $LiH_xD_{1-x}$ alloys [28]. The results obtained in these recent applications will be discussed in the next Section in conjuction with those deduced from the HVW calculations.

**2. Combining the HVW results with those derived from a thermodynamic model.**

The so called cbΩ model states that the defect Gibbs energy $g^i$ is given by

$$g^i = c^i B\Omega \qquad (1)$$

where $B$ is the isothermal bulk modulus, $\Omega$ the mean volume per atom and $c^i$ is a constant independent of temperature and pressure for a given solid; the superscript $i$ stands for the defect process under study, i.e., $i$ = formation (*f*), migration (*m*) and activation (*act*). By differentiating Eq.(1) in respect to pressure (*P*) and considering that the defect volume $\upsilon^i$ is given by

$$\upsilon^i = \left.\frac{dg^i}{dP}\right|_T \tag{2}$$

a combination of Eqs.(1) and (2) leads to:

$$\frac{\upsilon^i}{g^i} = \frac{\left.\frac{dB}{dP}\right|_T - 1}{B} \tag{3}$$

Since $g^i = h^i - Ts^i$, where $h^i$ and $s^i$ stand for the defect enthalpy and the defect entropy of the corresponding process, we may approximate $g^i \approx h^i$ in the low temperature range. Thus, Eq.(3) may be approximately written as

$$\frac{\upsilon^i}{h^i} \approx \frac{\left.\frac{dB}{dP}\right|_T - 1}{B} \tag{4}$$

The validity of Eqs(3) and (4) has been checked for various defect processes and/or categories of solids [29]. In addition, these equations have been used [30] to show that there exists an interconnection between the defect parameters and the stress induced electric signals in ionic crystals, which provides the basis for the explanation of the emission of precursory electric signals before rupture [31-38].

Hereafter, we focus on the application of these equations to LiH by making use of the single crystal synchrotron X-ray diffraction [39] measurements on solid $^7$LiH up to 366 GPa. Loubeyre et al reported that their *P* vs *V* data are reproduced very well through an equation of state which depends on three parameters, the volume at ambient pressure, the isothermal bulk modulus *B*=32.2 GPa and the pressure derivative $dB/dP = 3.53$. In particular, let us first discuss the Schottky defect formation process and apply Eq. (3) to the case of the Schottky pair formation volume $v^f$. In order to do that, we first need to estimate $g^f$, while only the formation enthalpy $h^f$ is experimentally available. Thus, the value of the Schottky defect formation entropy $s^f$ whose value can be calculated from the relation $s^f = (-dg^f/dT)\big|_P$ considering that $g^f$ is given from Eq (1) and that the quantities B and Ω depend on temperature. By following the procedure described in Refs. [20] and [24] and taking into account the (thermal) expansivity data and the $dB/dT$ – value reported at T=300 K by Gerlich and Smith [25], we find $s^f \approx 10$ $k_B$, where $k_B$ is the usual Boltzmann constant. By inserting this value as well as the experimental value of the formation enthalpy $h^f$ =2.33 eV into the well known relation $g^f = h^f - Ts^f$, we find $g^f \approx 2.07$ eV at T=300 K. By inserting these values of $g^f$, B and $dB/dP$ into Eq. (3), we find that the Schottky defect formation volume $v^f$ =15.67 cm$^3$/mole at T=300 K. This is larger from the molecular volume 10.25 cm$^3$/mole which reflects the following: The relaxation volume around the anion- and cation-vacancy is positive, i.e., the neighboring ions around the anion-and cation-vacancy are moving on the average outwards as found by HVW.

We now turn to the calculation of the volume $v^m$ for the migration of the negatively charged lithium vacancy. Following the same procedure as the one explained above for $s^f$, we find that the entropy $s^m$ for the negatively charged lithium vacancy migration has a value between 1 and 2 $k_B$. This, when recalling that the experimental value of the lithium vacancy migration enthalpy is $h^m$ =0.54 eV, and using the relation $g^m = h^m - Ts^m$, leads to $g^m$ = 0.50 eV. Then Eq. (3), when considering the aforementioned values of B and $dB/dP$, gives $v^m$ =3.8 cm$^3$/mole.

The aforementioned values of $\upsilon^f$ and $\upsilon^m$ indicate that the value of the activation volume $\upsilon^{act}$ (for the intrinsic region of the conductivity plot in which the activation enthalpy $h^{act}$ is equal to $\frac{h^f}{2}+h^m$) is $\upsilon^{act} = \frac{\upsilon^f}{2}+\upsilon^m$. Thus $\upsilon^{act}=11.6$ cm$^3$/mol. Recall that this $\upsilon^{act}$ value was deduced above upon using the experimental values [19] $h^f=2.33$ eV and $h^m=0.54$ eV and hence $h^{act} \approx =1.70$ eV. If we use the value $h^{act}=1.37$ eV obtained in the HVW first principles calculations –instead of the experimental value $h^{act} \approx =1.70$ eV- we find $\upsilon^{act} \approx 9.4$ cm$^3$/mole which is smaller by only around ~20% from the aforementioned value $\upsilon^{act}=11.6$ cm$^3$/mol. Both these $\upsilon^{act}$ values are positive, which in view of Eq.(2) dictate that $g^{act}$ increases upon increasing the hydrostatic pressure, thus the electrical conductivity (as well the self diffusion coefficients) of LiH decreases upon pressurizing.

### 3. Conclusion

We conclude by stating that the HVW first principles calculations as well as thermodynamical aspects based on the cBΩ model support the view that the ionic conductivity of LiH is likely to decrease upon increasing the hydrostatic pressure. Direct measurements verifying this expectation are still lacking, thus it is of high interest to carry out such experiments.